\documentclass[prl,aps,showpacs,twocolumn,floatfix]{revtex4}
\usepackage{epsfig} \usepackage{graphics} \usepackage{bm}
\usepackage{amssymb}
\begin{document}
\title{Cooper Pairs with Broken Time-Reversal, Parity, and Spin-Rotational 
Symmetries in Singlet Type-II Superconductors}
\author{O.~Dutta and A.G.~Lebed}
\affiliation{Department of Physics,
University of Arizona, 1118 E. 4th Street, Tucson, AZ 85721, USA}
\date{\today}

\begin{abstract}
We show that singlet superconductivity in the Abrikosov
vortex phase is absolutely unstable with respect to the appearance 
of a chiral triplet component of a superconducting order parameter.
This chiral component, $p_x + i p_y$, breaks time-reversal, parity,
and spin rotational symmetries of the internal order parameter, 
responsible for a relative motion of two electrons in the
Cooper pair.
We demonstrate that the symmetry breaking Pauli paramagnetic effects
can be tuned by a magnetic field strength and direction and can be
made of the order of unity in organic and high-temperature layered
superconductors.

\end{abstract}

\maketitle

Conventional superconductivity is characterized by pairs of electrons
with opposite spins, known as Cooper pairs.
In their relative coordinate system, the internal wave function of the
conventional Cooper pair [1] is isotropic with zero total spin and zero
orbital angular momentum.
Among modern materials, there are two types of unconventional
superconductors: singlet d-wave and triplet ones, where the latter
are characterized by broken parity symmetry of the internal Cooper
pairs wave function [2,3].
Singlet d-wave superconductivity has been firmly established in
quasi-two-dimensional (Q2D) high temperature [4] and organic [5]
materials.
On the other hand, heavy fermion [6,7], Sr$_2$RuO$_4$ [8],
ferromagnetic [9], and (TMTSF)$_2$X [10] compounds are candidates
for a triplet superconducting pairing.
Recently, it has been demonstrated [11] that a triplet component of the
internal order parameter is always generated in the Abrikosov vortex
phase of singlet superconductors due to the Pauli paramagnetic
spin-splitting effects.
Phenomenological theory of the singlet-triplet mixed order parameters
in the Abrikosov phase has been considered in Ref.[12].

In this context, the most important from physical point of view
symmetry of the internal superconducting order parameter
is a time-reversal one.
According to a general theory of unconventional superconductivity
[2,3], a time-reversal symmetry of the internal orbital order parameter
may be broken for multi-component order
parameters.
The corresponding chiral Cooper pairs  possess non-zero spontaneous
orbital magnetic momenta.
Experimentally, such situation is realized in super-fluid $^3$He, where
the so-called $A$- and $A_1$-phases are characterized by superfluid
Cooper pairs with non-zero magnetic orbital momenta.
A possibility of a chiral triplet order parameter, $p_x + i p_y$, to exist
in an unconventional superconductor Sr$_2$RuO$_4$ is widely
discussed [8], in particular, in a  connection with the recent remarkable
measurements of the Kerr effect [13].
Nevertheless, in our opinion, the chiral triplet order
parameter, $p_x + i p_y$, has not been  firmly established in Sr$_2$RuO$_4$ 
since it seems to contradict to some other experimental
data [14-16].

A purpose of our Letter is to show that a chiral triplet order parameter
always appears in singlet superconductors in the Abrikosov vortex
phase due to the Pauli spin-splitting paramagnetic
effects.
In some sense, this means that there are no singlet type-II superconductors.
Indeed, as shown below, the internal orbital wave function of the Cooper
pairs is always characterized by a singlet-chiral triplet mixed order
parameter, which breaks time-reversal, parity, and spin-rotational
symmetries.
As a result of the time-reversal symmetry breaking, the Abrikosov vortices
are shown to possess an unusual distribution of magnetization.
It is important that the symmetry breaking effects exist both for attractive
and repulsive effective electron interactions in a triplet
channel.

To the best of our knowledge, this fundamental phenomenon has been
overlooked in the past.
In particular, it was not considered in our Letter [11] due to a special (parallel) 
orientation of a magnetic field.
Though the suggested theory is applied to all type II superconductors,
below, we emphasize on Q$2$D d-wave organic and high-Tc  
superconductors.
In the former case, as shown, the symmetry breaking effects can be always 
made of the order of unity in an inclined magnetic 
field.

We start from a simplest generalization of the BCS Hamiltonian for the
case of unconventional superconductors [2,3],
\begin{eqnarray}\label{ham}
H  & = & \sum_{\vec{p}, \sigma}  \epsilon_{\sigma}(\vec{p}) a^{\dagger}_{\vec{p}, \sigma} a_{\vec{p}, \sigma}  \nonumber\\
& + &\frac{1}{2} \sum_{\vec{p}, \vec{p}', \vec{q}, \sigma} V( \vec{p},\vec{p}' )
a^{\dagger}_{\vec{p} + \vec{\frac{q}{2}}, \sigma} a^{\dagger}_{-\vec{p}+\vec{\frac{q}{2}}, -\sigma}
a_{-\vec{p}'+\vec{\frac{q}{2}}, -\sigma} a_{\vec{p}'+\vec{\frac{q}{2}}, \sigma} , 
\nonumber\\
\end{eqnarray}
where the effective electron interactions do not depend on electrons 
spins, $s = \sigma /2 \ (\sigma = \pm 1)$.
In Eq.(1), $2$D electron energy in a magnetic field is
$\epsilon_{\sigma}(\vec{p}) = (p^2_x+p^2_y)/2m - \sigma \mu_{B}H$,
where $\mu_B$ is the Bohr magneton;  $\vec{q}$ corresponds to motion of a 
center of mass of the Cooper pairs, $\vec{p}$ and $\vec{p}'$ correspond to 
relative motion of the electrons in
the Cooper pairs.

Below, we extend a classical method [17] to derive Ginzberg-Landau 
(GL) equations to the case of a singlet-triplet mixed order 
parameter.
In particular, we represent effective electron interactions potential as 
a sum of singlet and triplet parts,  
$ V_(\vec{p}, \vec{p}') = V_s (\vec{p}, \vec{p}')
+ V_t (\vec{p}, \vec{p}') $, and define the finite temperature
normal and Gor'kov Green functions [18],
\begin{eqnarray}
G_{\sigma, \sigma}(\vec{p},\vec{p}';\tau) &= &
- \langle T_{ \tau} a_{\sigma}(\vec{p}, \tau) a^{\dagger}_{\sigma}(\vec{p}', 0 ) \rangle, \nonumber\\
F_{\sigma,-\sigma}(\vec{p}, \vec{p}';\tau) &=& \langle T_{\tau} a_{\sigma}(\vec{p}, \tau)
a_{-\sigma}(-\vec{p}', 0 ) \rangle, \nonumber\\
F^{\dagger}_{\sigma,-\sigma}(\vec{p}, \vec{p}';\tau) &=& \langle
T_{\tau} a^{\dagger}_{\sigma}(-\vec{p}, \tau) a^{\dagger}_{-\sigma}(\vec{p}', 0 ) \rangle.
\end{eqnarray}

In this case, singlet and triplet order parameters can be defined by means of Gor'kov Green function as,
\begin{eqnarray}\label{orders}
\Delta_{s}(\vec{p}, \vec{q}) &=& -\frac{1}{2} \sum_{\vec{p}'} V_{s}(\vec{p}, \vec{p}') T \sum_{\omega_n} \nonumber\\
& & \left [ F_{+, -}(\vec{p}'+\frac{\vec{q}}{2}, \vec{p}'-\frac{\vec{q}}{2};
{\it i} \omega_n) - \right . \nonumber\\
& & \left . F_{-, +}(\vec{p}'+\frac{\vec{q}}{2}, \vec{p}'-\frac{\vec{q}}{2}; {\it i} \omega_n) \right ],
\end{eqnarray}
\begin{eqnarray}\label{ordert}
\Delta_{t}(\vec{p}, \vec{q}) &=& -\frac{1}{2} \sum_{\vec{p}'} V_{t}(\vec{p}, \vec{p}') T \sum_{\omega_n} \nonumber\\
& & \left [ F_{+, -}(\vec{p}'+\frac{\vec{q}}{2}, \vec{p}'-\frac{\vec{q}}{2}; {\it i} \omega_n) + \right . \nonumber\\
& & \left . F_{-, +}(\vec{p}'+\frac{\vec{q}}{2}, \vec{p}'-\frac{\vec{q}}{2}; {\it i} \omega_n) \right ],
\end{eqnarray}
where $\omega_n=(2n+1)\pi T$ is the Matsubara frequency [18].

In the Letter, we calculate superconducting transition temperature by 
means of the linearized Gor'kov Eqs. (2)-(4) for the following singlet 
and triplet parts 
of the effective electron interactions,
$V_s(\vec{p}, \vec{p}')=- 8 \pi g_s \cos(2 \phi) \cos(2 \phi')$ and
$V_t(\vec{p}, \vec{p}')=- 8 \pi g_t  \cos( \phi - \phi' )$, where
$\phi$($\phi'$) is an azimuthal angle corresponding to $2$D electron momentum
$\vec{p}$($\vec{p}'$). 

Below, we consider the case, where $d_{\rm x^2-y^2}$- superconducting
order parameter,
\begin{equation}\label{dp1}
\Delta_s(\vec{p}, \vec{q}) = \sqrt{2} \Delta_s(\vec{q}) \cos(2 \phi),
\end{equation}
corresponds to a ground state at $H=0$. 
Whereas a triplet component of the order parameter,
\begin{equation}\label{dp2}
\Delta_t(\vec{p}, \vec{q}) = \sqrt{2} [ \Delta^1_t(\vec{q}) \cos(\phi)
+ \Delta^2_t(\vec{q}) \sin(\phi)],
\end{equation}
is a secondary effect and appears only in the presence of 
a magnetic field.

     Solving Eqs.(2)-(6) at $T_c-T \ll T_c$, where $T_c$ is the
transition temperature  to singlet phase (\ref{dp1}) at $H=0$, 
we obtain
\begin{eqnarray}\label{gleq}
\frac{\Delta_s(\vec{q})}{g_s} &=& \it{A_1} \Delta_s(\vec{q})
+ \it{D_1} \Delta^1_t(\vec{q}) + \it{D_2} \Delta^2_t(\vec{q}), \nonumber\\
\frac{\Delta^1_t(\vec{q})}{g_t} &=& \it{A_2} \Delta^1_t(\vec{q}) +
\it{C} \Delta^2_t(\vec{q}) + \it{D_1} \Delta_s(\vec{q}), \nonumber\\
\frac{\Delta^2_t(\vec{q})}{g_t} &=& \it{C} \Delta^1_t(\vec{q}) +
\it{A_3} \Delta^2_t(\vec{q}) + \it{D_2} \Delta_s(\vec{q}),
\end{eqnarray}
where
\begin{eqnarray}\label{coeff}
\it{A_1} & = & \pi T \sum_{n\ge 0} \left [ \frac{2}{\omega_n} - \frac{\it{v_F}^2}{4 \omega^3_n}
 ( q^2_x + q^2_y ) \right ] , \nonumber\\
\it{A_2} &=& \pi T \sum_{n\ge 0} \left [ \frac{2}{\omega_n} - \frac{\it{v_F}^2}{4 \omega^3_n}
  \left ( \frac{3}{2}q^2_x + \frac{1}{2} q^2_y \right) \right ] , \nonumber\\
\it{A_3} &=& \pi T \sum_{n\ge 0} \left [ \frac{2}{\omega_n} - \frac{\it{v_F}^2}{4 \omega^3_n}
  \left ( \frac{1}{2} q^2_x + \frac{3}{2} q^2_y \right) \right ] , \nonumber\\
\it{C} &=& - \pi T  \biggl( \sum_{n\ge 0} \frac{1}{4 \omega^3_n}  \biggl) 
\frac{v^2_F(q_x q_y+ q_y q_x)}{2}, \nonumber\\
\it{D_1} &=& - \mu_B H ( \pi T)   \biggl( \sum_{n \ge 0} \frac{1}{\omega^3_n}  \biggl)
(v_F q_x), \nonumber\\
\it{D_2} &=& \mu_B H ( \pi T)  \biggl(  \sum_{n \ge 0} \frac{1}{\omega^3_n}  \biggl)
(v_F q_y) ,
\end{eqnarray}
with $v_F$ being the Fermi velocity.

Note that the principle difference between our Eqs.(\ref{gleq}), (\ref{coeff}) and the
results of Ref. [11] is that the singlet component (\ref{dp1}) is coupled to two
triplet components (\ref{dp2}), which, as shown below, results in a time reversal symmetry breaking.

    In the presence of a magnetic field, the gauge transformation,
$\vec{q} \rightarrow \vec{\Pi} \equiv - {\it i}\vec{\nabla} - (2e/c) \vec{A} $, where
$2{\it e}$ is charge of the Cooper pair, results in the following GL equations,
\begin{widetext}

\begin{eqnarray}
\label{gl1}
&& \left [ t - \xi^2_{\parallel} \left ( \Pi^2_x + \Pi^2_y  \right ) \right ] \Delta_s(x, y)
 - \sqrt{\frac{7\zeta(3)}{2}} \frac{\mu_B H}{\pi T} \xi_{\parallel}
\left [ \Pi_x \Delta^1_t(x,y) - \Pi_y \Delta^2_t(x,y) \right ] = 0,\\
\label{gl2}
&& \left [ 1 - \frac{g_t}{g_s} - \frac{\xi^2_{\parallel}}{2} \left ( 3 \Pi^2_x + \Pi^2_y \right )
 \right ] \Delta^1_t(x, y) - \frac{\xi^2_{\parallel}}{2} \left [ \Pi_x \Pi_y + \Pi_y \Pi_x \right ] \Delta^2_t(x,y)
 + g_t  \sqrt{\frac{7\zeta(3)}{2}} \frac{\mu_B H}{\pi T}
\xi_{\parallel}\Pi_x\Delta_s(x, y) = 0, \\
\label{gl3}
&& \left [ 1 - \frac{g_t}{g_s} - \frac{\xi^2_{\parallel}}{2} \left ( \Pi^2_x + 3 \Pi^2_y \right )
 \right ] \Delta^2_t(x, y) - \frac{\xi^2_{\parallel}}{2} \left [ \Pi_x \Pi_y + \Pi_y \Pi_x \right ] \Delta^1_t(x,y)
 - g_t \sqrt{\frac{7\zeta(3)}{2}} \frac{\mu_B H}{\pi T}
\xi_{\parallel}\Pi_y\Delta_s(x, y) = 0,
\end{eqnarray}
\end{widetext}
where we also perform the Fourier transformation with respect to $\vec{q}$. 
Note that, in
Eqs.~(\ref{gl1})-(\ref{gl3}), $g_s>g_t$ are effective electron coupling
 constants in singlet and triplet channels respectively, 
 $\xi_{\parallel} = \sqrt{7 \zeta(3)}
\it{v_F}/4 \sqrt{2} \pi T_c$ is in-plane GL coherence length,
and $t=(T_c-T)/T_c \ll 1$.
Eqs.~(\ref{gl1})-(\ref{gl3}) directly demonstrate instability of singlet
superconductivity with respect to a generation of two triplet
components (\ref{dp2}) since they do not have a solution for
$\Delta^1_t(x,y)=\Delta^2_t(x,y)=0$.

{\it High Tc Superconductors}: for $|g_t|<<g_s$, Eq.~\ref{gl1} transforms to the conventional equation to determine the superconducting nucleus
$[ t - \xi^2_{\parallel} \Pi^2 ] \Delta_s(x, y)=0$ with $\Pi^2=\Pi^2_x+\Pi^2_y$.
The GL Eqs. (10),(11) for two triplet order parameters (\ref{dp2}) simplify to
\begin{eqnarray}\label{neu}
& & \Delta^1_t(x, y) + g_t \sqrt{\frac{7\zeta(3)}{2}} \biggl( \frac{\mu_B H}{\pi T}\biggl)
\xi_{\parallel}\Pi_x\Delta_s(x, y) = 0 ,  \nonumber\\
& & \Delta^2_t(x, y) - g_t \sqrt{\frac{7\zeta(3)}{2}} \biggl( \frac{\mu_B H}{\pi T}
\biggl) \xi_{\parallel}\Pi_y\Delta_s(x, y) = 0 .
\end{eqnarray}
Here, we consider the case where a magnetic field
is applied perpendicular to the conducting planes of a high-Tc superconductor. Then
the upper critical field is given by the conventional formula
$H^{\bot}_{c_2}= t \phi_0/2 \pi \xi^2_{\parallel}$. 
For magnetic fields $H \leq H_{c_2}$
in gauge $\vec{A}=(0,Hx,0)$, the order parameter of the superconducting nucleus
is given by
\begin{equation} \label{gla}
\bf \left[
\begin{array}{c}
 \Delta_s(x)   \\
                      \\
 \Delta^1_t(x)   \\
                  \\
 \Delta^2_t(x)   \\
\end{array}
\right] = \bf \left[
\begin{array}{c}
 \exp \left (-\frac{t x^2}{2 \xi^2_{\parallel}} \right )  \\
 - i g_t \sqrt{t} \alpha(H) \left[ \frac{\sqrt{t} x}{\xi_{\parallel}} \right ]
\exp \left (- \frac{t x^2}{2\xi^2_{\parallel}} \right )  \\
- g_t \sqrt{t} \alpha(| H |) \left[ \frac{\sqrt{t} x}{\xi_{\parallel}} \right ]
\exp \left (- \frac{t x^2}{2\xi^2_{\parallel}} \right )   \\
\end{array}
\right],
\end{equation}
where $\alpha(H)=\sqrt{7\xi(3)/2}(\mu_B H/\pi T_c)$. Note that the recent measurements
of the upper critical field in high-Tc superconductors [19] give
$H_{c_2} \sim H_P \sim T_c/\mu_B$, which means that the effects of the singlet-triplet mixing (13) can be made of the order of unity
if $|g_t| \sim g_s$.

       It is important that the chiral triplet component of the order
parameter (\ref{gla}) is associated with angular momentum,
\begin{equation}
L =  sgn(H) g^2_t \alpha^2(H_{c_2})
\left [ \frac{(T_c-T) x}{T_c \xi_{\parallel}} \right ]^2
\exp \left [- \frac{(T_c-T) x^2}{T_c  \xi^2_{\parallel}} \right ], 
\end{equation}
which is directed along the applied magnetic field and possesses a non-trivial coordinate dependence.
It is instructive to rewrite superconducting order parameter (\ref{dp1}), (\ref{dp2}), (\ref{gla})
in a form where its spin structure and chirality are shown explicitly,
\begin{eqnarray}\label{or2}
&&\Delta(x,y;p)=\Delta_s(x,y) \cos(2 \phi)*\left (
|\uparrow \downarrow > - |\downarrow \uparrow > \right )  \nonumber\\
& & + i \frac {\Delta_{t}(x,y)}{2} [\ sgn(H) \ p_x+ i p_y] 
*\left (|\uparrow \downarrow > + |\downarrow \uparrow > \right ),
\end{eqnarray}
where $p_x=\cos(\phi)$ and $p_y=\sin(\phi)$.

         The presence of both singlet and triplet components in Eq.~(\ref{or2}) breaks
parity and spin-rotational symmetries of the internal order parameter, whereas
the chiral triplet component, $p_x+ip_y$, breaks its time-reversal 
symmetry.
The appearance of the chiral component, $p_x+ip_y$, results in the counter 
clockwise relative motion of the two electrons in the Cooper 
pair. 
This leads to the appearance of orbital magnetic moment of the Cooper pair, 
applied exactly opposite to the direction of the external magnetic 
field. 
It is important that coordinate dependence of the above mentioned magnetic 
moment can be expressed through singlet superconducting gap, 
$\Delta_s(x,y) = |\Delta_s(x,y)| \exp[i \phi(x,y)]$, 
in the Abrikosov vortex phase 
in the following way,
\begin{equation} \label{abr}
M \sim - |\Delta_s(x,y)|
\left [ \left ( \frac{\partial |\Delta_s(x,y)|}{\partial y} \right ) v_x -
 \left ( \frac{\partial |\Delta_s(x,y)|}{\partial x} \right ) v_y \right ] ,
\end{equation}
where $v_x = \frac{1}{2m} [ \frac{\partial \phi(x,y)}{\partial x} - \frac{2e}{c} A_x]$ 
and $v_x = \frac{1}{2m} [ \frac{\partial \phi(x,y)}{\partial y} - \frac{2e}{c} A_y]$ are 
the corresponding components of the superfluid
velocity.
We propose to measure the spatial distribution of the magnetic moment
(\ref{abr}), which is different from the spatial distribution of a magnetic moment due to
the Meissner currents, to prove the symmetry breaking effect suggested 
in the Letter.

     {\it Layered Organic Superconductors}: in a typical Q2D organic material
[5], the upper critical field perpendicular to the conducting layers, $H^{\bot}_{c_2} \ll H_P$, whereas the parallel upper critical field, $H^{\|}_{c_2} \gg H_P$, where $H_P$ 
is the Clogston paramagnetic limit [3].
Under such conditions, we suggest experiments in an inclined magnetic field, where 
only perpendicular component of the field is important. 
In this case, all equations derived above are still valid if we replace $H$ by its perpendicular component, $H \rightarrow H \sin\theta$, where $\theta$ is the 
angle between a magnetic field and the conducting layers. 
An analysis of a such experiment shows that the suggested symmetry breaking 
effects are maximal (i.e., of the order of unity)
at $H\sin \theta \sim H_P$. 
Therefore, we expect that angular dependence of the upper
critical field has to demonstrate deviations from the so-called "effective mass" model
in the vicinity of some small angle $\theta^{*}\sim H^{\bot}_{c_2}/H_P \ll 1$ 
(see Fig.1).
We propose detailed measurements of the upper critical fields in organic superconductors to detect possible deviations from the "effective mass" model 
in order to prove the existence of symmetry breaking effects suggested 
in this Letter.

%------------------------------
\begin{figure}[t]
\begin{center}
\epsfig{file=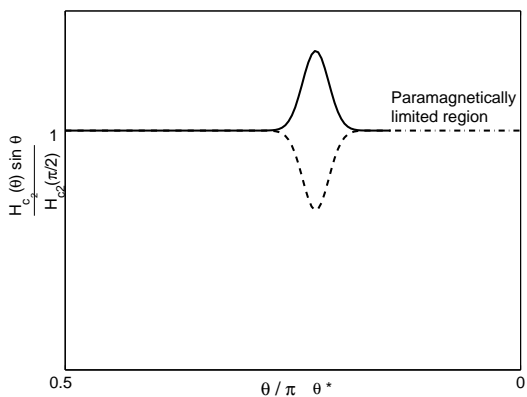,width=7cm} \caption{\label{fig1} 
Schematic diagram of the suggested
angular dependence of the upper critical field for a highly anisotropic
Q2D organic superconductor. 
The dash dotted
line denotes the region, where the Pauli paramagnetic effect destroy superconductivity.
Due to the singlet-triplet mixing effects the normalized upper critical field
is higher (lower) than $1$ depending
on a sign of the effective electron interactions in a triplet channel. }
\end{center}
\end{figure}
%----------------------------

In conclusion, we point out that the phenomenon, suggested in the
Letter, is different from the singlet-triplet mixing effects in non-centrosymmetric
superconductors [20-22,3,12]. 
Indeed, the so-called Lifshitz invariant [3], responsible for the singlet-triplet
mixing effects [20-22,12], does not exist in zero magnetic field in an arbitrary
case.
In the Latter, we show that it always appears in the Abrikosov vortex phase
in any singlet type-II superconductor due to the Pauli spin-splitting 
effects [11].
Other words, the main message of the Letter is that the singlet-triplet mixing
effects, which break time-reversal, parity, and spin-rotational symmetries of 
the internal order parameter, appear in any singlet type-II 
superconductor.
In Q2D organic and high-T$_c$ superconductors, these effects are
expected to be of the order of unity.

 One of us (A.G.L.) is thankful to N.N. Bagmet, V.V. Kabanov, and Z. Tesanovic 
 for useful
discussions.


\begin{thebibliography}{99}
\bibitem{leb-2}
The triplet chiral order parameter, $p_x + i p_y$, seems not to be in 
agreement with the existence of the upper paramagnetic-like limiting field [15] as 
well as with the existence of lines of nodes on a superconducting gap
[16] in Sr$_2$RuO$_4$.
\end{thebibliography}
\end{document}